# Maximizing bipolar sensitivity for anomalous Nernst thermopiles in heat flux sensing in amorphous GdCo alloys

Miho Odagiri, Hiroto Imaeda, Ahmet Yagmur, Yuichiro Kurokawa, Satoshi Sumi, Hiroyuki Awano, and Kenji Tanabe*

*tanabe@toyota-ti.ac.jp

**Abstract**

A Heat Flux Sensor (HFS) facilitates the visualization of heat flow, unlike a temperature sensor, and is anticipated to be a key technology in managing waste heat. Recently, an HFS utilizing the Anomalous Nernst Effect (ANE) has been proposed garnering significant interest in enhancing the transverse Seebeck coefficient. However, ideal materials for HFS not only require a large transverse Seebeck coefficient but also meet several criteria including low thermal conductivity and a bipolar nature of the transverse Seebeck coefficient, especially a negative coefficient. In this study, we have investigated ANE in amorphous ferrimagnetic GdCo alloys, revealing their numerous advantages as HFS materials. These include a large transverse Seebeck coefficient, extremely low thermal conductivity, large negative sensitivity, unparalleled bipolar sensitivity, versatility for deposition on various substrates, and a small longitudinal Seebeck coefficient. These qualities position GdCo films as promising candidates for the advancement of HFS technology.

**Introduction**

    A Heat Flux Sensor (HFS) is pivotal in visualizing heat flow, a capability beyond the scope of standard temperature sensors, and is deemed essential in the management of waste heat. Traditional HFS systems, based on the Seebeck effect where the electromotive force is generated parallel to a temperature gradient, are notably expensive due to complex fabrication processes and the need for precise sensitivity calibration.

    In a significant development, Zhou et al. have proposed an HFS design utilizing the Anomalous Nernst Effect (ANE). This approach generates an electromotive force perpendicular to the temperature gradient, presenting several advantages[1]. The use of thin film technology in these ANE-based HFS devices promises ease of manufacturing, improved reproducibility, and the potential for a flexible sensor. The ANE is mathematically represented as:

$$\boldsymbol{E} = S_{\text{ANE}}\left(\frac{\boldsymbol{M}}{|\boldsymbol{M}|} \times (-\boldsymbol{\nabla}T)\right), \qquad (1)$$

where $\boldsymbol{E}$ is an electric field induced by ANE, $S_{\text{ANE}}$ is a transverse Seebeck coefficient, $\boldsymbol{\nabla}T$ is the temperature gradient, and $\boldsymbol{M}$ is magnetization. $S_{\text{ANE}}$ is a critical factor in ANE research. Historically, the ANE has been studied as one of the transport properties from the viewpoint of fundamental physics since the Nernst effect had been discovered in 1887[2]. A great turning point in ANE research is the studies by Sakuraba et al. from the viewpoint of applied physics[1,3-4]. They proposed the application into the thermoelectric power source and HFS using a transverse thermopile structure in Fig. 1. After that many groups have reported the studies on the ANE at room temperature[5-11]. Another breakthrough in ANE research was the discovery by Ikhlas et al. in 2017 of a relatively large $S_{\text{ANE}}$ value of 0.6 µVK$^{-1}$ in antiferromagnetic $Mn_3Sn$[12]. This finding was remarkable because the enhancement in $S_{\text{ANE}}$ was attributed to the increased Berry curvature at Weyl points near the Fermi energy. This discovery spurred further research, with several groups identifying Weyl magnetic materials exhibiting even larger $S_{\text{ANE}}$ values[13-27]. For example, Sakai et al. and Guin et al. observed a giant ANE in the full-Heusler ferromagnet $Co_2MnGa$, achieving an $S_{\text{ANE}}$ of 6 µVK$^{-1}$ at room temperature[13,14].

    The key parameter in evaluating an HFS is the ratio of thermal electromotive force ($V$) to the heat flux density ($j$), expressed as a sensor sensitivity $V/j$. This parameter is, however, a quantitative variable and depends on the sensor area and its shape. Hence, $V/j$ is unsuitable as a parameter in evaluating materials. Instead of $V/j$, we use the ratio of electric field ($E$) induced by ANE to the heat flux density ($j$), expressed as $E/j$,

to evaluate materials for ANE-based HFS. $E/j$ is an intensifying variable and independent of the sensor area and its shape. According to Fourier's law and the definition of the transverse Seebeck coefficient ($S_{ANE}$), this ratio is given by:

$$\frac{E}{j} = \frac{S_{ANE}}{\kappa}, \qquad (2)$$

where $\kappa$ is the thermal conductivity. This equation underlines the necessity for materials with a large transverse Seebeck coefficient and low thermal conductivity for effective HFS design. When Zhou et al. proposed an innovative ANE-type HFS, they also demonstrated its viability using FeAl alloys[1]. They showed that single-crystalline $Fe_{81}Al_{19}$ alloy films on MgO substrates and polycrystalline films on amorphous substrates both exhibit large $S_{ANE}$ values (+3.4 and +3.1 μVK$^{-1}$, respectively), indicating that the sensitivity is largely independent of the substrate type. They fabricated an HFS that consists of the $Fe_{81}Al_{19}$-Au thermopile on a Si substrate and its sensor sensitivity $V/j$ is +0.020 μVW$^{-1}$m$^2$ when the sensor area is $1 \times 1$ cm$^2$. Note that the sensor sensitivity differs from the sensitivity for materials and depends on the sensor area and shape of the thermopile. From this sensor sensitivity, the material sensitivity $E/j$ for these films is approximately estimated to be +0.20 μmA$^{-1}$. The unit of $E/j$ is (Vm$^{-1}$)(Wm$^{-2}$)$^{-1}$ and it is simplified as mA$^{-1}$. Uchida et al. fabricated an HFS using a thermopile structure of $Co_2MnGa$ and Au with a giant $S_{ANE}$ of +6 μVK$^{-1}$ at room temperature and its sensor sensitivity $V/j$ is +0.110 μVW$^{-1}$m$^2$ under a magnetic field of 1 kOe[28]. In contrast, the sensitivity at zero field reduces to +0.016 μVW$^{-1}$m$^2$ and they pointed out the importance of the remanent magnetization. Modak et al. focused on SmCo alloys, known for their high coercive force, and found that their sensitivity remains stable even at zero magnetic field[29]. Tanaka et al. highlighted the importance of using two materials with similar longitudinal Seebeck coefficients ($S_{xx}$) to minimize the Seebeck effect's influence[30]. The criteria for two materials of the HFS is not only large $S_{ANE}$, but also low thermal conductivity, bipolarity of $S_{ANE}$, similar small $S_{xx}$, and the flexibility to fabricate onto any substrates.

Rare-earth (RE) and transition-metal (TM) alloys, such as GdFeCo, are notable for their ferrimagnetic properties, characterized by the antiparallel alignment of the magnetic moments of TM and RE elements[31-34]. These materials exhibit a diverse range of magnetic behaviors depending on their composition. In RE-rich films, the net magnetization aligns with the magnetic moment of the rare-earth metal, while in TM-rich films, the transition metal predominantly contributes to the net magnetization. The boundary composition between these RE-rich and TM-rich alloys is known as the magnetization compensation point (MCP). Additionally, the amorphous nature of these alloys, arising from their lack of crystal structure, results in relatively low thermal conductivities. For instance, $Gd_{21}Fe_{72}Co_7$ and $Tb_{21}Fe_{73}Co_6$ films exhibit thermal conductivities of approximately 5.5 and 4.6 $Wm^{-1}K^{-1}$, respectively, at room temperature[35]. The amorphous nature is advantageous for various practical applications, as it enables the deposition of these alloys on a wide range of substrates at room temperature. The unique combination of properties of RE-TM alloys, including their magnetic behavior, low thermal conductivity, and flexible deposition, make them valuable for practical use. Notably, they have already found applications in commercial magneto-optical storage systems, demonstrating their utility and versatility in advanced technological applications.

Several groups have reported the ANE in RE-TM alloys[7,10,31,33,36-39]. Seki et al. explored the anomalous Ettingshausen effect, a counterpart to the ANE, in GdCo alloys and estimated the transverse Seebeck coefficient to be 0.18 $\mu VK^{-1}$ in $Gd_{22}Co_{78}$[36]. Liu et al. studied the ANE in $Gd_{16}Co_{84}$ and $Gd_{26}Co_{74}$ alloys[37], discovering that the polarity of the ANE signal is dictated by the magnetization orientation of the Co sub-lattices rather than the overall net magnetization of GdCo. Odagiri et al. studied the composition dependence of the ANE in TbCo alloys and discovered the coexistence of a large transverse Seebeck coefficient of nearly 1.0 $\mu VK^{-1}$ and a large coercive force of more than 1 kOe[39]. However, the material sensitivity for an ANE-based HFS has never been investigated in the RE-TM alloys yet.

In this study, we present a comprehensive analysis of the composition dependence of the ANE in GdCo alloy films. We found that doping Gd into GdCo alloys substantially enhances the sensitivity $(E/j)$, achieving a value of 0.23 $\mu mA^{-1}$ near the MCP which is comparable to $Co_2MnGa$. The origin of high sensitivity is the relatively large $S_{ANE}$ and extremely small $\kappa$. Moreover, GdCo exhibits bipolarity of the sensitivity $E/j$. This study, therefore, highlights the potential of GdCo alloys in advancing technologies that leverage the ANE.

**Results and Discussion**
Anomalous Nernst effect

The transverse Seebeck coefficient, $S_{ANE}$, was measured under conditions of a perpendicular magnetic field and an in-plane thermal gradient using the Hall bar samples as shown in Figs. 2(a-b). In Fig. 2(c), we plot the detected voltage as a function of an external magnetic field under the application of a constant thermal gradient for a GdCo sample with $x$ = 15.5 at room temperature. The ANE voltage $V$ is defined as half the voltage difference at high magnetic fields between positive and negative saturations.

We observe that $V$ increases with a rising temperature difference. The absence of a hysteresis curve in this plot suggests that the sample possesses in-plane magnetic anisotropy. Figure 2(d) illustrates the ANE electric field $E$ as a function of the temperature gradient $\nabla T$, derived from the data in Fig. 2(c). $E$ shows a direct proportionality to $\nabla T$. The blue dotted line represents a linear fitting, and its slope corresponds to the transverse Seebeck coefficient $S_{ANE}$.

Figure 2(e) illustrates the composition-dependent behavior of $S_{ANE}$ in GdCo films. The behavior is perfectly consistent with TbCo results[39], suggesting that the characteristics of $S_{ANE}$ are related not to an orbital momentum in 4f electron of rare earth elements but to random Co position in the amorphous alloy. A notable change in the sign of $S_{ANE}$ is observed between compositions $x$ = 15.5 and 21.7, corresponding to the change from a TM-rich to an RE-rich alloy composition. It means that ANE in GdCo is predominantly influenced by the Co moment. Unlike other materials with large $S_{ANE}$ values such as $Fe_{81}Al_{19}$[1], which exhibit positive polarity, RE-rich GdCo alloys demonstrate significant negative polarity. This characteristic is crucial for applications requiring both positive and negative $S_{ANE}$ values, such as a HFS. Additionally, the magnitude of $S_{ANE}$ near the MCP remains relatively stable, around 1.0 µVK$^{-1}$. The presence of a large $S_{ANE}$ around the MCP, coupled with the sign change at the MCP, is vital for the development of HFS modules. The observed large $|S_{ANE}|$ of around 1.0 µVK$^{-1}$ is inconsistent with the previous reports by Seki et al. and Liu et al. They used multilayer structures like Al(4 nm)/GdCo(30 nm)/Al(4 nm) and Ta(3 nm)/GdCo(4.7, 6.2 nm)/Pt(3 nm)/Ta(1 nm), potentially leading to an underestimation of $|S_{ANE}|$ due to the shunting effect of the metallic layers.

In the specific case of $x$ = 21.7, close to the MCP, the decrease in the magnitude of $S_{ANE}$, reminiscent of the decrease observed in $Tb_{19.9}Co_{80.1}$[39], is observed in Fig. 2(e). This decrease originates from the coexistence of TM-rich and RE-rich regions like $Tb_{19.9}Co_{80.1}$.

Thermo-electric transport properties

We investigated the composition dependencies of electrical resistivity $\rho_{xx}$, Hall resistivity $\rho_{yx}$, and longitudinal Seebeck coefficient $S_{xx}$ in GdCo alloys to elucidate the factors contributing to the observed large $S_{ANE}$ values. Figures 3(a-c) show the schematic diagram of the measurement setup for $\rho_{xx}$, $\rho_{yx}$, and $S_{xx}$, and Figures 3(d-f) shows $\rho_{xx}$, $\rho_{yx}$, and $S_{xx}$ as a function of the GdCo composition. As depicted in Fig. 3(f), $S_{xx}$ decreases with increased Gd doping and becomes notably small around the MCP. According to the theories on thermo-electric transport properties, $S_{ANE}$ can be decomposed into two distinct contributions,
$$S_{\text{ANE}} = \alpha_{yx}\rho_{xx} + \alpha_{xx}\rho_{yx}.$$
Here, $\alpha_{xx}$ is the longitudinal thermoelectric constant, and $\alpha_{yx}$ is the transverse thermoelectric constant. The terms $\alpha_{yx}\rho_{xx}$ and $\alpha_{xx}\rho_{yx}$ correspond to $S_1$ and $S_2$, respectively. All parameters can be determined from $\rho_{xx}$, $\rho_{yx}$, $S_{xx}$ and $S_{\text{ANE}}$. $S_1$ represents an intrinsic mechanism where the transverse electric current is generated directly from the temperature gradient, independent of the Seebeck effect. Figure 3(g) shows $S_{\text{ANE}}$, $S_1$, and $S_2$ as a function of the GdCo composition. Our findings indicate that $S_1$ is significantly larger than $S_2$, highlighting the importance of enhancing the transverse thermoelectric constant $\alpha_{yx}$ and the resistivity $\rho_{xx}$ in enhancing $S_{\text{ANE}}$. The magnitude of $\alpha_{yx}$ is dramatically enhanced with slight Gd doping and slowly decreases with Gd doping as depicted in Figs. 3(h-i). $|\alpha_{yx}|$ is relatively large even as $x = 30$. The enhancement of $S_{\text{ANE}}$ originates from the increase in $\rho_{xx}$ and in $\alpha_{yx}$.

The origin of the large $|\alpha_{xy}|$ is commonly attributed to Berry curvature, an effective magnetic field in momentum space. However, the application of Berry curvature to amorphous materials is challenging due to the absence of periodic crystal structure. Despite this, recent studies have demonstrated the influence of Berry curvature on the anomalous Hall effect[40] and the ANE[25] in amorphous alloys like FeGe and FeSn. Following a widely accepted approach to validate the contributions of Berry curvature to the ANE[41-42]. we have plotted $|\alpha_{yx}|$ against $|\sigma_{yx}|$ in Figs. 4(a-b). By assuming a high-temperature limit, we obtain a ratio of $|\alpha_{yx}/\sigma_{yx}|\sim k_B/e$, where $k_B$ is the Boltzmann constant and $e$ is the elementary charge. The ratio $|\alpha_{yx}/\sigma_{yx}|$ for the GdCo films is found to be as same as $k_B/2e$, mirroring those observed in topological magnets as shown in Fig. 4(a). This consistent pattern in $|\alpha_{yx}/\sigma_{yx}|$ ratios underlines the influence of the Berry curvature on the anomalous Hall effect and ANE in GdCo films. When comparing with the crystalline $GdCo_5$, which shares the same crystal structure as the $SmCo_5$ alloy, the $|\alpha_{yx}/\sigma_{yx}|$ values for amorphous GdCo films are slightly lower than those calculated for $GdCo_5$ using first-principles calculations[47]. This observation is reminiscent of the comparison between amorphous FeSn alloys and their crystalline counterparts, $Fe_3Sn$ and $Fe_3Sn_2$, suggesting that the short-range order present in $GdCo_5$ also persists even in the amorphous GdCo alloys.

Sensitivity in evaluating materials for ANE-based HFS

We examined the composition dependence of the sensitivity, the ANE electric field per heat flux density. Figure 5(a) shows the schematic diagram of the measurement setup for the sensitivity. In Fig. 5(b), the detected voltage is plotted as a function of an external magnetic field for $x = 11.1$ at room temperature. Note that the voltage at zero-magnetic field is as same as that under the high magnetic field because the GdCo films have in-plane magnetic anisotropy. Figure 5(c) illustrates the relationship between the ANE electric field $E$ and the heat flux density $j$. $E$ is completely proportional to $j$ and its slope corresponds to the sensitivity of the material for an HFS application. Figure 5(d) shows the composition-dependent behavior of the sensitivity $E/j$ in GdCo films. The dependency of $E/j$ is reminiscent of that of $S_{ANE}$. For anisotropic materials, $S_{ANE}$ depends on the directions of the external magnetic field and temperature gradient. However, in isotropic materials like amorphous, polycrystalline, and cubic materials, $S_{ANE}$ is setup-independent. The similarity between $E/j$ and $S_{ANE}$ is reasonable owing to their linear relation ($E/j = S_{ANE}/\kappa$). In contrast, the remarked increase appears with Gd doping in the low doping region, compared with $S_{ANE}$. The increase is related to the suppression of the thermal conductivity. The magnitude of $E/j$ peaks at 0.23 µmA$^{-1}$ at $x = 23.7$, comparable to Co$_2$MnGa[28], as shown in Figure 5(e). Although $S_{ANE}$ in GdCo($x = 23.7$) is much smaller than 6 µVK$^{-1}$ in Co$_2$MnGa, its $E/j$ value is about the same as that of Co$_2$MnGa (0.22 µmA$^{-1}$) owing to the high thermal conductivity of 23 Wm$^{-1}$K$^{-1}$ in bulk Co$_2$MnGa[12].

Thermal conductivity in ultrathin films

Here, we consider thermal conductivity in the GdCo films. There is no method for measuring thermal conductivity in an ultrathin metallic film in the direction perpendicular to the film. We assume the GdCo film is isotropic, indicating $S_{ANE}$ is as same as the transverse Seebeck coefficient under the application of the in-plane magnetic field and the temperature gradient perpendicular to the films. $\kappa$ is derived by $S_{ANE}/(E/j)$. Figure 5(f) depicts the composition dependence of the thermal conductivity $\kappa$. The thermal conductivity significantly decreases with Gd doping and reaches 5 Wm$^{-1}$K$^{-1}$, which is quite similar to 4.6-5.5 Wm$^{-1}$K$^{-1}$ in Gd$_{21}$Fe$_{72}$Co$_7$ and Tb$_{21}$Fe$_{73}$Co$_6$ films in the previous report[35]. Thermal conductivity is comprised of contributions from conduction electrons, phonons, and magnons. Given the typically negligible contribution of magnon thermal conductivity, we focus on the other two. Applying the Wiedemann–Franz law and assuming that the resistivity along the z-direction, $\rho_{zz}$, is the same as $\rho_{xx}$, we calculate the electron thermal conductivity, $\kappa_{ele}$, using the equation:

$$\kappa_{ele} = LT\sigma_{zz},$$

where $L$ is Lorentz number (2.44×10$^{-8}$ WΩK$^{-2}$), $T$ is the temperature, and $\sigma_{zz} = \rho_{xx}^{-1}$. Since the total thermal conductivity $\kappa$ consists of $\kappa_{ele}$ and the phonon thermal conductivity $\kappa_{\text{pho}}$, we determined $\kappa_{\text{pho}}$ as $\kappa - \kappa_{\text{ele}}$ as shown in Figure 5(g). The significant reduction in $\kappa$ aligns closely with $\kappa_{\text{ele}}$, revealing that the remarked increase in the sensitivity $E/j$ in the low doping region is closely related to the enhanced resistivity $\rho_{xx}$ in Fig. 3(a). Meanwhile, the phonon thermal conductivity remains sufficiently low due to the amorphous nature of the materials. Therefore, enhancing the resistivity is crucial to obtaining higher sensitivity in GdCo.

## Comparison of other materials

Finally, let us compare the characteristics of GdCo with other materials. The GdCo alloys have the sensitivity $|E/j|$ of 0.20 μmA$^{-1}$ at a wide composition range around the MCP. Although this value is comparable to 0.22 μmA$^{-1}$ in Co$_2$MnGa[28], 0.20 μmA$^{-1}$ in Fe$_{81}$Al$_{19}$[1], and 0.18 μmA$^{-1}$ in Sm$_{20}$Co$_{80}$[29]. A distinguishing feature of GdCo alloys is their bipolarity in $E/j$, which allows for a thermopile structure composed exclusively of GdCo pairs. Figure 6(a) shows the summary of the reported $E/j$, revealing only the Re-rich GdCo has negative giant sensitivity. Co$_2$MnGa, Fe$_{81}$Al$_{19}$, Sm$_{20}$Co$_{80}$, and Fe$_{79}$Ga$_{21}$ exhibit only positive $E/j$ polarity and therefore require a different material to complete the thermopile structure. Consequently, the bipolar sensitivity of such thermopile structures, defined as $(E/j)_+ - (E/j)_-$, is effectively halved compared to those in our study, as illustrated in Figure 6(b), where $(E/j)_{+(-)}$ represents $E/j$ in materials with $S_{\text{ANE}} > 0 (< 0)$. For HFS applications, which necessitate two materials with similar $S_{xx}$ but larger $S_{\text{ANE}}$, the properties of GdCo are particularly advantageous. While Tanaka et al. achieved this using Fe$_{79}$Ga$_{21}$ and Fe$_{79}$Ga$_{21}$/Ni$_{10}$Cu$_{90}$[30], GdCo alloys offer a simpler solution by utilizing TM-rich and RE-rich compositions near the MCP. Additionally, GdCo alloys are versatile in terms of substrate compatibility; they can be deposited on any substrate at room temperature, enhancing their practical applicability. Furthermore, the transport properties of GdCo are remarkably similar to those in TbCo. Doping Tb into GdCo alloys could potentially induce a suitable magnetic anisotropy, adding another functionality to these materials.

Our findings give a guideline to obtain higher sensitivity in GdCo alloys, indicating the importance of an increase in resistivity. Since $S_{\text{ANE}}$ is mainly attributed to the intrinsic conversion, $S_{\text{ANE}}$ is roughly proportional to $\rho_{xx}$. Additionally, $\kappa$ is suppressed by the increase in resistivity because the phonon thermal conductivity remains sufficiently low due to the amorphous nature. Therefore, the sensitivity $E/j$ is roughly represented as $(\alpha_{yx}/LT)\rho_{xx}^2$ and increases with the square of the resistivity. Doping slight impurity elements into GdCo alloys may exhibit a significant enhancement of the sensitivity.

Conclusion

Our comprehensive study on the ANE in ferrimagnetic GdCo alloy films has yielded significant insights. Our results reveal that doping Gd into the alloy significantly enhances the sensitivity. Remarkably, the magnitude of the sensitivity peaks at 0.20 µmA$^{-1}$ over a broad composition range, including the MCP, where a notable change in polarity is observed. The maximum value of 0.23 µmA$^{-1}$ at $x = 23.7$ is comparable to the highest sensitivity found in Co$_2$MnGa. Additionally, the observed bipolarity of the high sensitivity is particularly advantageous for a thermopile structure. Twice sensitivity can be obtained by using the TM- and Re-rich materials, compared with the previous studies. The composition dependence of $S_{\mathrm{ANE}}$ suggests that the high sensitivity originates from the coexistence of a relatively large $S_{\mathrm{ANE}}$ (over 1.0 µVK$^{-1}$) and low thermal conductivity (less than 5 Wm$^{-1}$K$^{-1}$) across a wide composition range, including the MCP. Moreover, our comprehensive analysis suggests that the large $S_{\mathrm{ANE}}$ is attributed to Berry curvature despite its amorphous nature. GdCo alloys demonstrate an ability to be deposited on any substrate and possess a small longitudinal Seebeck coefficient. These are attributed to underscore the potential of GdCo films for use in ANE-based HFS applications.

**Methods**

Sample preparation

The sample structure is Si$_3$N$_4$(10 nm) /Gd$_x$Co$_{100-x}$(20 nm) /Si$_3$N$_4$(3 nm), which was deposited on both two SiO$_2$ glass substrates and a Si substrate with a thermally oxidized layer using magnetron rf and dc sputtering techniques at room temperature. The base pressure in the chamber is less than 1x10$^{-5}$ Pa. The Si$_3$N$_4$ layers act as a protection layer from the oxidation of GdCo films. For one of the glass samples, a Hall bar structure was formed using a metal mask, as depicted in Fig. 2(a). For the other glass samples and the Si-substrate sample, the films were not patterned. The GdCo alloys, varying in Gd composition from 0 to 35, were created through a co-sputtering method with Gd and Co cathodes. Gd(99.9%) and Co(99.9%) targets were purchased from Chemiston Co. The compositions of these alloys were verified via energy-dispersive X-ray spectroscopy.

Measurement method for thermal-electric transport properties

The transverse Seebeck coefficient, $S_{\mathrm{ANE}}$, was measured under conditions of a perpendicular magnetic field and an in-plane thermal gradient, using the Hall bar samples as shown in Fig. 2(a). The temperature gradient was controlled using a heater and monitored with a T-type thermocouple (copper-constantan) attached to the sample holders. While there was a minor difference between the thermocouple-measured temperature gradient and the actual gradient within the sample, this discrepancy was calibrated using thermography. To ensure the accuracy of our measurement setup, we also measured the transverse Seebeck coefficient in a polycrystalline Py film produced by magnetron sputtering, obtaining a value of +0.53 μVK$^{-1}$, consistent with previous findings.[48] The polarity of $S_{ANE}$, as defined in Fig. 2(a), aligns with established conventions: $S_{ANE}$ is positive (>0) in materials such as Co, Ni, and Py, and negative (<0) in Fe. Additionally, the longitudinal Seebeck coefficient, electrical resistivity, and Hall resistivity were measured using the same sample, with the setups illustrated in Figs. 3(a-c). The four-terminal method was employed for electrical resistivity measurements.

Measurement method for sensitivity

The sensitivity of a material for an HFS, $E/j$, was measured under the condition of an in-plane magnetic field using the unpatterned samples prepared on glass substrates as shown in Fig. 5(a). The sample was sandwiched between Cu blocks and heat was applied to the sample in the direction perpendicular to the film by a heater. The heat flow through the sample is monitored by a commercialized HFS connected to the sample in series. To reduce a contact thermal resistance, we used thermal greases (8.3 Wm$^{-1}$K$^{-1}$) for a CPU device in a PC. To calibrate the accuracy of our measurement setup, we also measured $E/j$ in a Py film produced by magnetron sputtering, obtaining values of +0.017 - +0.018 μmA$^{-1}$, consistent with $S_{ANE}/\kappa$ in the previous reports[48-49]. The polarity of $E/j$ aligns with established conventions: $E/j$ is positive (negative) as $S_{ANE} > 0$ (< 0).

**Data Availability**

The data that support the findings of this study are available from the corresponding author upon reasonable request.

**Acknowledgements**

This work was mainly supported by the Paloma Environmental Technology Development Foundation.

**Author information**

Authors and Affiliations

Toyota Technological Institute, Nagoya, 468-8511, Japan

M. Odagiri, H. Imaeda, A. Yagmur, S. Sumi, H. Awano, K. Tanabe

University of Leeds, Leeds, LS2 9JT, UK

A. Yagmur

Graduate School and Faculty of Information Science and Electrical Engineering, Kyushu University, Fukuoka, 819-0395, Japan

Y. Kurokawa

Contributions

K.T. conceived the project. M.O. prepared the samples used in this study. M.O. mainly performed measurements and H.I., A.Y., Y.K., S.S., H.A., and K.T. supported her. M.O. and K.T. carried out the analysis. K.T. wrote the paper. All authors discussed the results and reviewed the paper.

Corresponding authors

Correspondence to Kenji Tanabe.

## Ethics declarations

Competing interests

The authors declare no competing financial or non-financial interests.

**Caption**

**Figure 1**

Schematic diagram of a transverse thermopile device based on ANE. The blue and orange wires indicate magnetic materials with positive and negative $S_{\mathrm{ANE}}$, respectively.

**Figure 2**

(a-b) Schematic illustrations of setups for measuring $S_{\mathrm{ANE}}$. (c) ANE voltage as functions of a magnetic field and temperature difference as $x = 15.5$ at room temperature. (d) Relationship between the electric field and temperature gradient as $x = 15.5$. The blue line indicates a linear fitting line and its slope corresponds to $S_{\mathrm{ANE}}$. (e) Transverse Seebeck coefficient as a function of GdCo composition. The blue triangle data were referred from the previous report[39]. The green dotted curves are guides to the eye.

**Figure 3**

(a-c) Schematic illustrations of setups for measuring electrical resistivity(a), Hall resistivity(b), and longitudinal Seebeck coefficient(c). (d-f) $\rho_{xx}$(d), $\rho_{yx}$(e), and $S_{xx}$(f) as a function of GdCo composition at room temperature. All the parameters were measured by using the same sample. The four-terminal method was used in the resistance measurement. The blue triangle data were referred from the previous report[39]. (g) $S_{\mathrm{ANE}}$, $S_1$, and $S_2$ as a function of GdCo composition. $S_2$ is derived by $S_{xx}\rho_{yx}/\rho_{xx}$ from Figs. 3(d-f). $S_1$ is derived by $S_{\mathrm{ANE}} - S_2$. (h-i) $\alpha_{yx}$(h) and its magnitude(i) as a function of GdCo composition.

**Figure 4**

(a) Summary of the relationship between $|\alpha_{yx}|$ and $|\sigma_{xy}|$ in topological magnets. $\sigma_{xy} = \rho_{yx}/(\rho_{xx}^2 + \rho_{yx}^2)$. La$_{0.3}$Sr$_{0.7}$CoO$_3$, Mn$_3$Sn, Mn$_3$Ge, Fe$_3$Sn$_2$, Fe$_{0.7}$Ga$_{0.3}$, Fe$_3$Sn, Co$_3$Sn$_2$S$_2$, and Co$_2$MnGa indicate La$_{0.3}$Sr$_{0.7}$CoO$_3$ cryst-bulk[46], Mn$_3$Sn cryst-bulk[12], Mn$_3$Ge cryst-bulk[47], Fe$_3$Sn$_2$ cryst-bulk[48], Fe$_{0.7}$Ga$_{0.3}$ cryst-film[16], Fe$_3$Sn cryst-bulk[24], Co$_3$Sn$_2$S$_2$ cryst-bulk[45, 49], and Co$_2$MnGa cryst-bulk[13], respectively. The blue triangle data were referred from the previous report[39]. (b) Comparison of the relationship between $|\alpha_{yx}|$ and $|\sigma_{yx}|$ in GdCo alloys with single-crystalline GdCo$_5$ alloy. The data in GdCo$_5$ was referred from the supplementary information in the previous report[47].

**Figure 5**

(a) Schematic illustrations of setups for measuring $E/j$. The directions of a magnetic field and heat flux in this setup differ from those in Figs. 2 (c-d). (b) ANE voltage as functions of a magnetic field and a heat flux density as $x = 11.1$ at room temperature. (c) ANE electric field as a function of a heat flux density as $x = 11.1$. The blue line indicates a linear fitting line. (d-e) $E/j$(d) and its magnitude(e) as a function of GdCo composition, *x*. (f) Thermal conductivity as a function of GdCo composition, *x*. The thermal conductivity $\kappa$ is derived by $S_{\text{ANE}}/(E/j)$. Here, we assumed that $S_{\text{ANE}}$ is isotropic. The blue dotted line corresponds to the thermal conductivity in $Tb_{21}Fe_{73}Co_6$[35]. (g) Total, electron, and phonon thermal conductivities as a function of GdCo composition, *x*. electron thermal conductivity is derived by $LT\sigma_{xx}$, where $L$ is the Lorentz number and $\sigma_{xx}$ is the longitudinal electrical conductivity along the in-plane direction. Here, we assumed that $\sigma_{xx}$ is isotropic. The phonon thermal conductivity is the difference between the total and electron thermal conductivities.

**Figure 6**

(a) Summary of sensitivity of heat flux densities. FeAl, SmCo, FeGa, and Co2MnGa indicate a $Fe_{81}Al_{19}$ film[1], a $Co_2MnGa$ film[27], an $Sm_{20}Co_{80}$ film[28], and a $Fe_{79}Ga_{21}$ film[29], respectively. GdCo(TM) and GdCo(RE) correspond to a $Gd_{15.5}Co_{84.5}$ film and a $Gd_{23.7}Co_{76.3}$ film in this study, respectively. Although $E/j$ is not shown in previous reports[1, 27-29], it was estimated from the reported sensor sensitivity $V/j$ and the sample shape, (b) Summary of sensitivity of thermopile for HFS. FeAl/Au, Co2MnGa/Au, SmCo/Au, and FeGa/CuNi indicate a $Fe_{81}Al_{19}$-Au thermopile[1], a $Co_2MnGa$-Au thermopile[27], a $Sm_{20}Co_{80}$-Au thermopile[28], and a $Fe_{81}Ga_{19}$-$Fe_{81}Ga_{19}/Cu_{90}Ni_{10}$ thermopile[29], respectively. GdCo/GoCo corresponds to the case of a $Gd_{15.5}Co_{84.5}$-$Gd_{23.7}Co_{76.3}$ thermopile device.

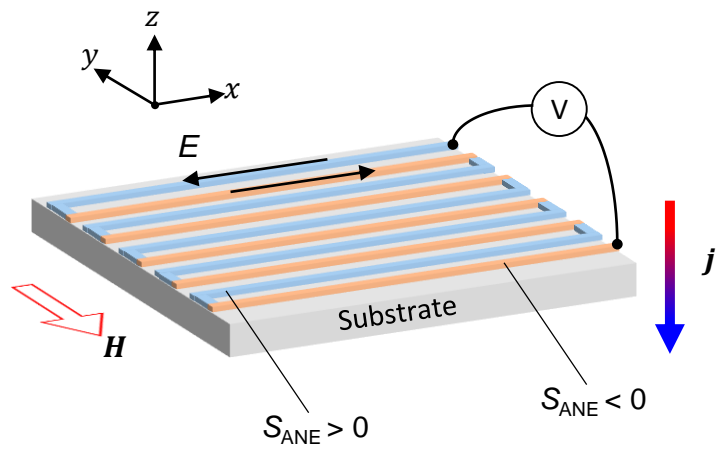

Figure 1

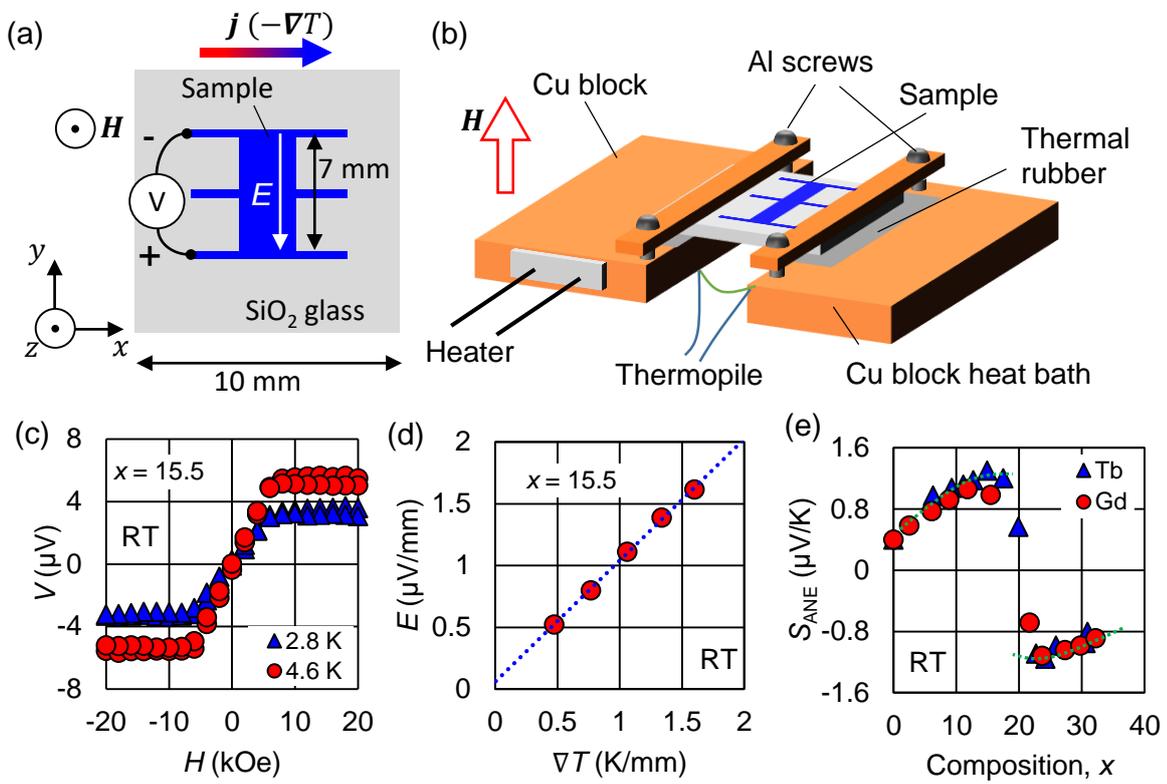

Figure 2

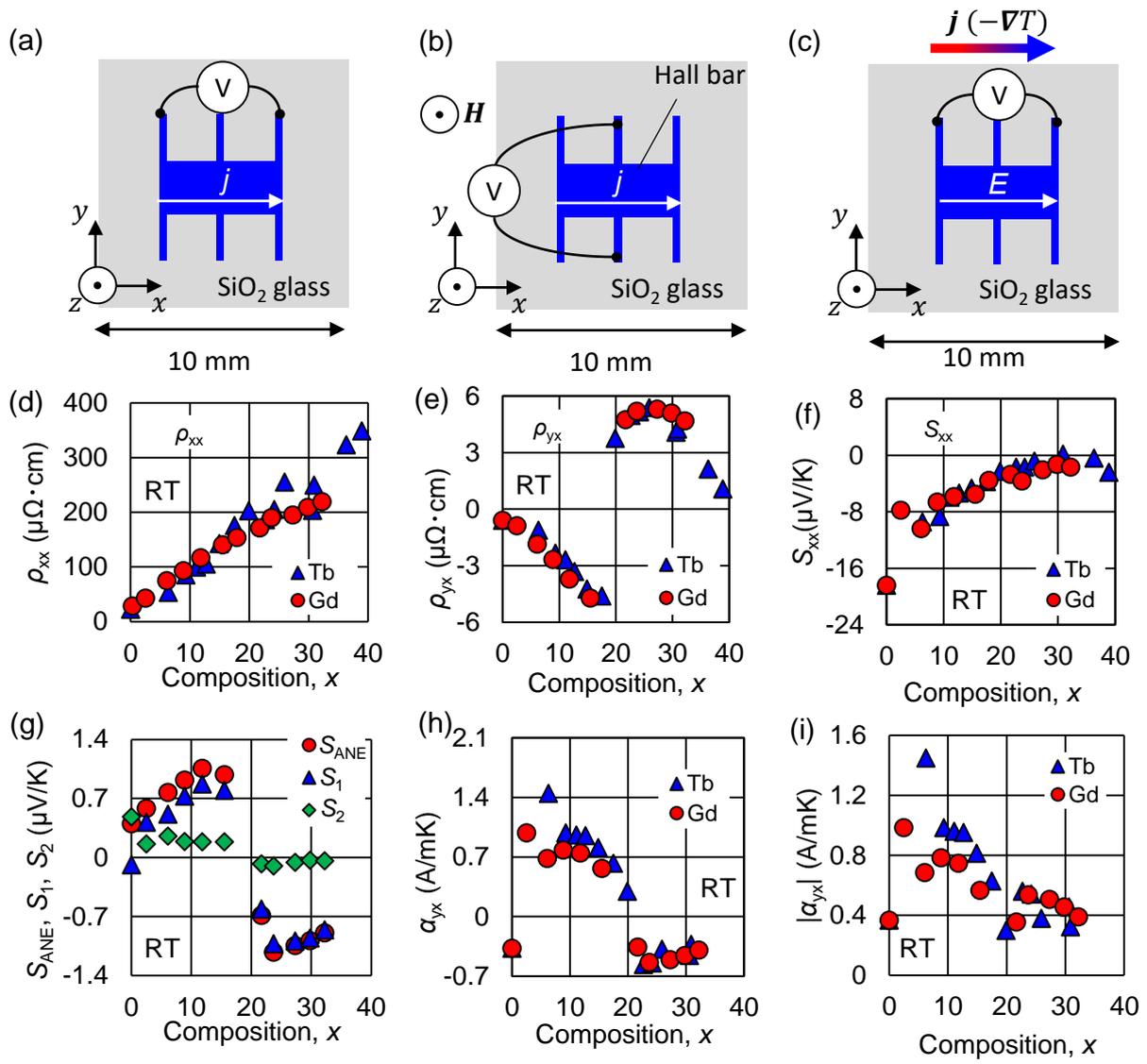

Figure 3

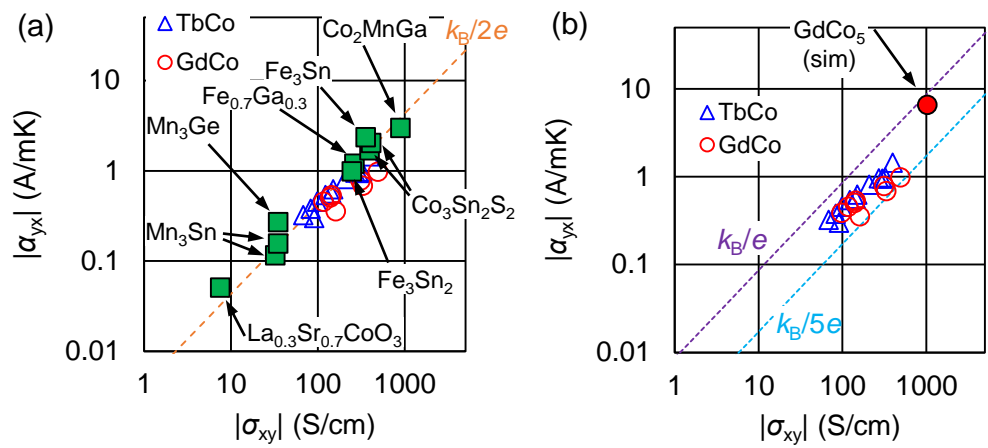

Figure 4

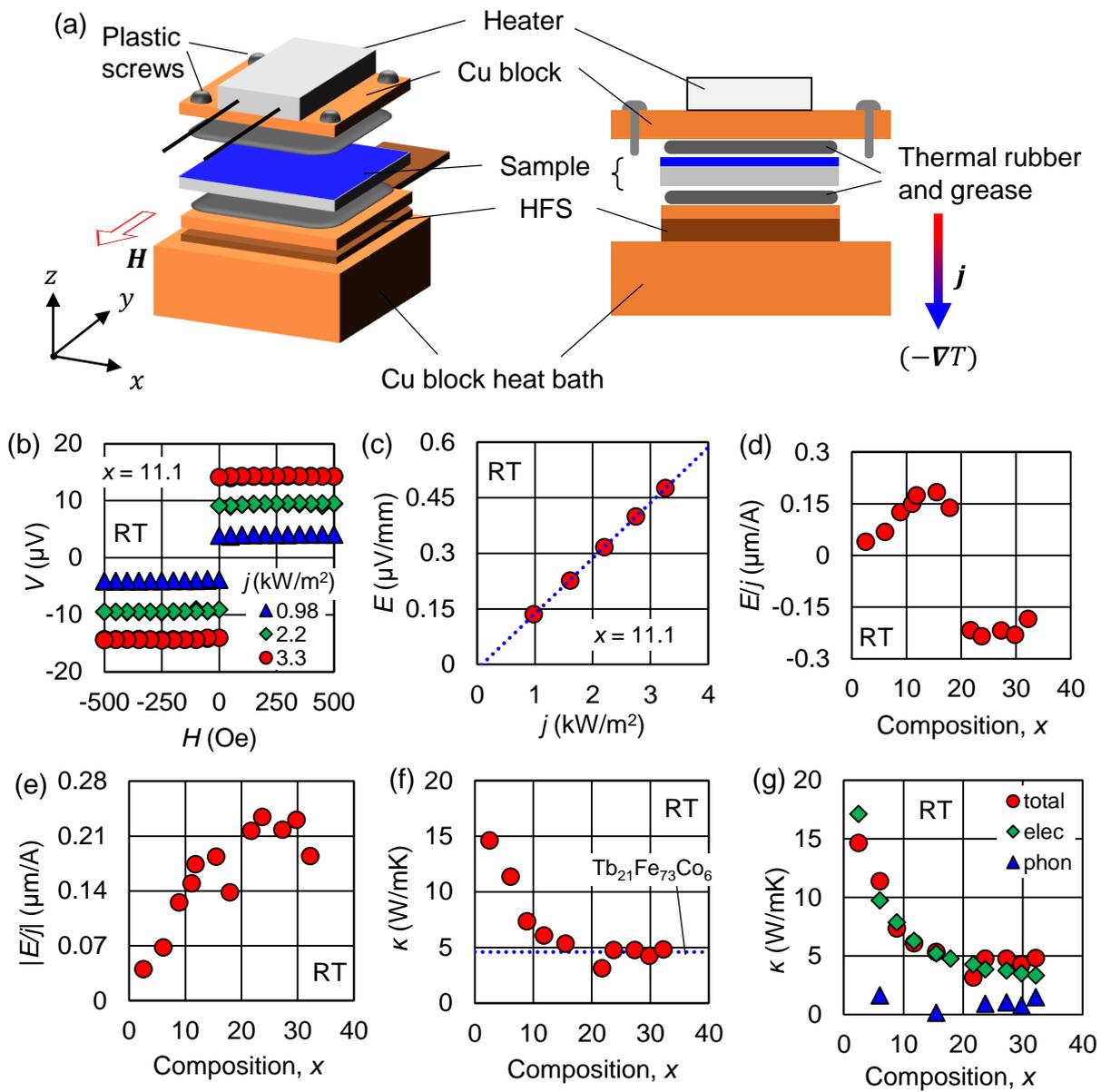

Figure 5

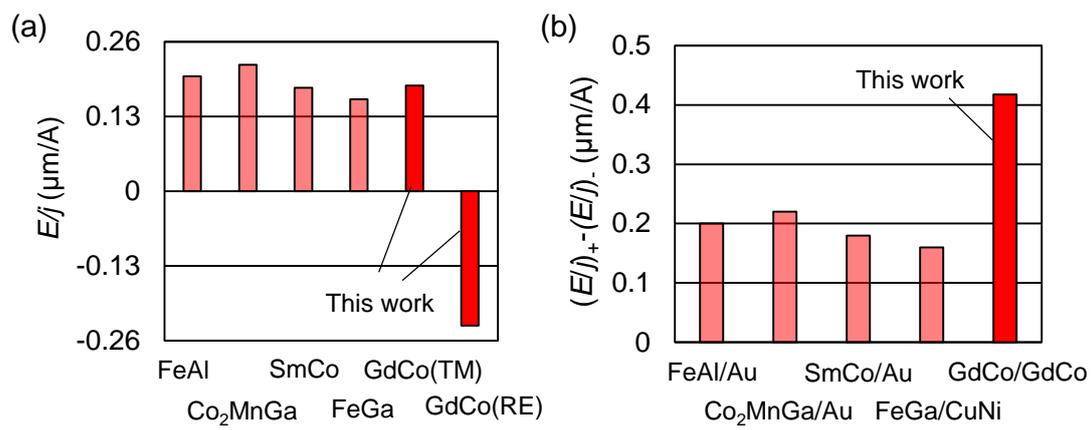

Figure 6